\documentclass[aps,prb,preprint,groupedaddress,floatfix]{revtex4}
\usepackage{latexsym}
\usepackage{amsmath}
\usepackage{amsfonts}
\usepackage{amssymb}
\usepackage{bm}
\usepackage{graphicx}
\usepackage{amsbsy}

\begin{document}
\def\deg{^\circ}
\newcommand{\ind}[1]{_{\mbox{\tiny#1}}}     
\newcommand{\supind}[1]{^{\mbox{\tiny#1}}}  
\newcommand{\unit}[1]{\, \mbox{#1}}               
\newcommand{\bs}[1]{\boldsymbol{\mathit{#1}}}
\newcommand{\lla}{\left\langle}
\newcommand{\rra}{\right\rangle}
\newcommand{\dst}{\displaystyle}

\graphicspath{{plots/}}

\pagestyle{empty}

\title{Stress Tensors of Multiparticle Collision Dynamics Fluids}
\author{Roland G. Winkler}
\email{r.winkler@fz-juelich.de}%
\author{Chien-Cheng Huang}%
\email{c.c.huang@fz-juelich.de}%
\affiliation{Institut f\"ur Festk\"orperforschung,
Forschungszentrum J\"ulich, D-52425 J\"ulich, Germany}
\date{\today}

\begin{abstract}
Stress tensors are derived for the multiparticle collision
dynamics algorithm, a particle-based mesoscale simulation method
for fluctuating fluids, resembling those of atomistic or molecular
systems. Systems with periodic boundary conditions as well as
fluids confined in a slit are considered. For every case, two
equivalent expressions for the tensor are provided, the internal
stress tensor, which involves all degrees of freedom of a system,
and the external stress, which only includes the interactions with
the confining surfaces. In addition, stress tensors for a system
with embedded particles are determined. Based on the derived
stress tensors, analytical expressions are calculated for the
shear viscosity. Simulations illustrate the difference in
fluctuations between the various derived expressions and yield
very good agreement between the numerical results and the
analytically derived expression for the viscosity.

\end{abstract}

\maketitle

\section{INTRODUCTION}

Soft matter systems, such as colloidal suspensions or polymer and
biopolymer solutions possess a wide range of length and time
scales. The need to bride the length- and time-scale gaps for
studies of these systems requires a simplified and coarse-grained
description of the solvent degrees of freedom. Several mesoscale
simulations techniques have been developed to meet this goal,
which adequately reproduce fluid behavior. Among them, the
multiparticle collision dynamics (MPC) method, originally proposed
by Malevanets and Kapral, \cite{male:99,male:00} has attracted
considerable attention over the last few years. In a wide spectrum
of applications, it has been shown that MPC reproduces fluid
properties adequately and accounts for hydrodynamic interactions,
as illustrated in the recent review articles Refs.
\onlinecite{kapr:08,gomp:08_p}.

Traditionally, there is a fundamental interest in the transport
properties of complex fluids. The coarse-grained simulation
approaches provide access to hydrodynamic phenomena on the
mesoscale. It has been shown that MPC is very well suited to study
non-equilibrium, rheological, and viscoelastic properties of such
fluids.\cite{padd:04,kiku:05,nogu:05,ripo:06,ryde:06,tao:08,cana:08,fran:08}
To fully characterize the equilibrium and non-equilibrium physical
properties of the fluid system, adequate microscopic
expressions---such as the stress tensor---have to be provided in
order to establish a link between the simulation degrees of
freedom and the macroscopic material properties, e.g., the
viscosity. Particular expressions for the stress tensor of an MPC
fluid have been provided in Refs.
\onlinecite{ihle:04,ihle:05,gomp:08_p} for a periodic system and
in Ref. \onlinecite{tao:08} for a slit geometry. Analytical
expressions for its viscosity have been derived by various
approaches.\cite{ihle:01,male:00,kiku:03,ihle:05,pool:05,nogu:08,kapr:08,gomp:08_p}

In this article, we will provide stress tensors at equilibrium and
under shear flow for an MPC fluid as well as for a system with
embedded point-like particles, which resembles the virial
formulation of molecular systems.
\cite{gree:47,irvi:50,beck:67,alle:87,wink:92,wink:93,davi:96,wink:02}
This formulation allows for a straightforward calculation of the
stress and provides an expression for the solvent-solute coupling.
Three dimensional systems with periodic boundary conditions are
considered as well as fluids in a slit geometry, which requires an
adaptation of the stress tensor due to wall interactions. Two
equivalent formulations of the stress tensor are provided in every
case, \cite{wink:92,wink:93,wink:02} corresponding either to the
mechanical definition of stress as force per area or as momentum
flux across a hypothetical plane.\cite{bird:87} The instantaneous
values of the respective expressions are different, but their
averages are identical. Since the provided expressions are novel,
we derive the shear viscosity from them, for both, a three
dimensional periodic system as well as a system confined between
walls under shear. We propose a modification of the MPC algorithm
in the presence of walls with respect to the inclusion of
wall-phantom particles. Compared to the original
algorithm,\cite{lamu:01} our formulation prevents any surface slip
of fluid particles.

The paper is organized as follows. In Sec. II, the multiparticle
collision dynamics method is described as well as its coupling to
a solute composed of mass points. In addition, the simulation
parameters are listed. Stress tensors are determined for MPC
fluids with periodic boundary conditions and those confined
between two parallel walls without external field in Sec. III. The
stress tensors in presence of shear flow for the same boundary
conditions are calculated in Sec. IV. In Sec. V, analytical
expressions for the viscosity are determined exploiting the
derived stress tensors. Section VI summarizes our findings.
Additional aspects of the fluid confined between surfaces are
discussed in Appendix A, namely the center-of-mass velocity in a
surface cell, and the wall collisional and wall kinetic stress
tensors.

\section{The Model}

\subsection{Multiparticle Collision Dynamics}

In the MPC algorithm, a fluid is represented by point particles of
mass $m$, which interact with each other by a stochastic process.
The algorithm consists of alternating streaming and collision
steps.\cite{male:99,kapr:08,gomp:08_p} In the streaming step, the
$N_s$ particle move ballistically and their positions change
according to
\begin{align} \label{mpc_ballistic}
{\bm r}_i(t) = {\bm r}_i(t-h) + h {\bm v}_i(t-h),
\end{align}
$i=1,\ldots,N_s$, in the time interval $h$, which we denote as
collision time. In the collision step, particles are sorted into
cubic cells of side length $a$ and their relative velocities with
respect to the center-of-mass velocity of every cell are rotated
around a randomly oriented axis by a fix angle $\alpha$. This
imposed stochastic process represents the effect of many real
collisions. In a collision step, mass, momentum and energy are
conserved which leads to the build up of correlations between the
particles and gives rise to hydrodynamic interactions. Hence, the
velocity of a particle changes according to
\begin{align} \label{mpc_collision}
{\bm v}_i(t) = \hat{\bm v}_i(t) + ({\mathbf  R}(\alpha)- {\mathbf E})
(\hat{\bm v}_i(t) - {\bm v}_{cm}(t)),
\end{align}
where $\hat{\bm v}_i(t)$ is the velocity before the collision,
${\mathbf R}(\alpha)$ is the rotation matrix, \cite{alla:02} ${\bm
v}_{cm} = \sum_{j}^{N_c} \hat{\bm v}_{j=1}/N_c$ is the
center-of-mass velocity of the particles contained in the cell of
particle $i$, and $N_c$ is the total number of fluid particles in
that cell. $\mathbf E$ is the unit matrix. Hence, the change of
momentum in a collision is
\begin{align} \label{mpc_momentum}
\Delta{\bm p}_i(t)& = m ({\bm v}_i(t)-\hat{\bm v}_i(t)) \\ \nonumber
& =  m({\mathbf R}(\alpha)- {\mathbf E}) [\hat{\bm v}_i(t) - {\bm v}_{cm}(t)].
\end{align}
Without external field, $\hat{\bm v}_i(t+h) ={\bm v}_i(t)$. In the
presence of such a field, however, the velocity may change during
the streaming step. Depending on the external field, additional
forces have to be included in Eq.
(\ref{mpc_ballistic}).\cite{lamu:01,cana:08,fran:08} To insure
Galilean invariance, a random shift is performed at any collision
step.\cite{ihle:01} Various alternative schemes for the stochastic
process have been proposed by now.\cite{alla:02,nogu:08} However,
the actual collision process is not important for the derivation
of a stress tensor, but it affects the dependence of the viscosity
on the MPC parameters.

\subsection{Solute dynamics, solvent-solute coupling}

In complex fluids, solute particles are embedded in the MPC
solvent. Here, we will assume that the solute is composed of mass
points, e.g., polymers,\cite{muss:05} which interact with each
other by pairwise potentials and their dynamics is treated by
molecular dynamics simulations (MD). More complex objects, such as
vesicles or solid bodies can also be embedded and their dynamics
be coupled to the fluid.\cite{gomp:08_p} We will consider $N_p$
objects (polymers) each composed of $N_m$ particles.  The
equations of motion of particle $k$ with mass $M_k$ of object
$\nu$ read
\begin{align} \label{eq_mot_sol}
M_k \ddot {\bm r}_k^{\nu} &  =  {\bm F}_k^{\nu} ,
\end{align}
where ${\bm F}_k^{\nu}$ is the total force. These equations are
solved by, e.g., a velocity Verlet algorithm,\cite{swop:82} which
provides the positions and velocities $\hat{\bm v}_k^{\nu}(t)$
starting at a time $t-h$.

The solute particles can easily be coupled to the solvent by
incorporating them in the collision step.\cite{male:00_1,muss:05}
For a collision cell with $N_c$ fluid particles and $N_m^c$ solute
particles, which may belog to different objects, the
center-of-mass velocity is given by
\begin{align} \label{md_mpc_coupling}
{\bm v}_{cm} (t) = \frac{\dst \sum_{i=1}^{N_c} m \hat{\bm v}_i(t) +
\sum_{\nu} \sum_{k}^{N_m^c} M_k \hat{\bm v}_k^{\nu}(t)}{\dst N_c m +
\sum_{k=1}^{N_m^c}M_k} ,
\end{align}
which yields the momentum change (\ref{mpc_collision})
\begin{align} \label{mpc_momentum_solute}
\Delta{\bm p}_k^{\nu}(t)=  M_k({\mathbf R}(\alpha)- {\mathbf E}) [\hat{\bm v}_k^{\nu}(t) - {\bm v}_{cm}(t)].
\end{align}
Here, $\nu$ and $k$ belong to those polymers and monomers,
respectively, which are within the considered collision cell. This
results in an exchange of momentum between the solvent and solute
degrees of freedom. The new monomer velocities are then used as
initial conditions for the MD simulation of the embedded
particles. Typically several MD steps are performed between
multiparticle collisions, because the applied force fields require
an integration time step, which is typically smaller than the
collision time.

\subsection{Simulation parameters}

In a simulation, a cubic system is considered with linear
extension $L$ and an average number of $N_c=10$ particles in a
collision cell. The rotation angle is set to $\alpha = 130 \deg$.
Length and time are scaled according to $\tilde r_{\beta} =
r_{\beta}/a$ and $\tilde t = t \sqrt{k_BT/ma^2}$, which
corresponds to the choice $k_BT =1$, $m=1$, and $a=1$, where $T$
is the temperature and $k_B$ the Boltzmann constant. The collision
time $\tilde h=0.1$ is applied, which is well in the collision
dominated regime of the fluid dynamics.\cite{ripo:04,ripo:05} In
the calculation of the viscosity, shear is imposed either by
Lees-Edwards boundary conditions \cite{alle:87} or by the opposite
movement of the confining parallel walls, with the shear rate
$\dot \gamma = 10^{-2} \sqrt{k_B T/ma^2}$.

\section{Stress tensor: No external field }

The actual form of the stress tensor depends on the boundary
conditions of the fluid. Here, we will address periodic boundary
conditions and solid walls. In general, the equation of motion of
the $\alpha$th ($\alpha \in \{x,y,z\}$) spatial component of the
$i$th  mass point is (${\bm r} = (r_{ix}, r_{iy},r_{iz})^T$)
\begin{align} \label{eq_motion_com}
m_i \ddot {r}_{i \alpha} = F_{i \alpha}.
\end{align}
In case of periodic boundary conditions, ${\bm r}_i$ referrers to
the position of the particle in the infinite system, i.e., we do
not jump to an image, which is located in the primary box, when a
particle crosses a boundary of the periodic lattice. Hence, ${\bm
r}_i$ is a continuous faction of time. Multiplication of Eq.
(\ref{eq_motion_com}) by $r_{i\beta}$ and summation over all $N_s$
particles yields
\begin{align} \label{eq_motion_vir}
\frac{d}{dt} \sum_{i=1}^{N_s} m_i v_{i \alpha}  r_{i \beta}=
\sum_{i=1}^{N_s} m_i v_{i \alpha} v_{i \beta} +
\sum_{i=1}^{N_s} F_{i \alpha} r_{i \beta}.
\end{align}
The average over time (or an ensemble) yields
\begin{align} \label{eq_motion_vir_avr}
\lla \sum_{i=1}^{N_s} m_i v_{i \alpha} v_{i \beta} \rra +
\lla \sum_{i=1}^{N_s} F_{i \alpha} r_{i \beta} \rra = 0,
\end{align}
because the term on the left hand side of Eq.
(\ref{eq_motion_vir}) vanishes for a diffusive or confined
system.\cite{wink:92,wink:93} Equation (\ref{eq_motion_vir_avr})
will be the basis for the derivation of stress tensors.

We will exploit the mechanical definition of the stress tensor
given by $\sigma_{\alpha \beta} = F_{\alpha}/A_{\beta}$, where
$F_{\alpha}$ denotes the total force  in the spatial direction
$\alpha$ across the surface of area $A_{\beta}$ with normal in the
spatial direction $\beta$.

\subsection{Periodic boundary conditions}

For a system with periodic boundary conditions, we assume that
initially all fluid particles are in the same box of the periodic
system, which we will denote as primary box. In the course of
time, the particles will diffuse out of that box. Some of them may
reenter and leave again several times. The periodic images of
particle $i$ are located at the positions ${\bm r}_i +{\bm R}_{\bm
n}$, with the lattice vectors
\begin{align} \label{lattice_vec}
{\bm R}_{\bm n} = (n_{x} L_{x}, n_{y} L_{y}, n_{z} L_{z})^T ,
\end{align}
corresponding to the lattice of images of the primary box. The
$n_{\alpha}$s are integer numbers and $L_{\alpha}$ denotes the box
length along the $\alpha$-direction. For a cubic lattice, $L = L_x
= L_y = L_z = \sqrt[3]{V}$, with $V$ the volume of the system.

The potential energy of the solute particles comprises inter- and
intramolecular pairwise contributions
\begin{align} \nonumber
U(\{{\bf r} \}) & = \frac{1}{2} \sum_{\nu=1}^{N_p}\sum_{\mu=1}^{N_p} \sum_{k=1}^{N_m}
\sum_{l=1}^{N_m} \sum_{{\bm n}} U_{kl}({\bm r}_k^{{\nu}}-{\bm r}_l^{{\mu}}-{\bm R}_n)
\\ \label{pot_energy} & + \frac{1}{2} \sum_{\nu=1}^{N_p} \sum_{k=1}^{N_m}
\sum_{l=1}^{N_m} U_{kl}^{\nu}({\bm r}_k^{{\nu}}-{\bm r}_l^{{\nu}}),
\end{align}
where the interaction of a particle with itself (not necessarily
its image) is excluded and $U^{\nu}$ includes all intramolecular
potentials of the object $\nu$. The sum over ${\bm n}$ accounts
for all the images of a particular particle. The total force ${\bm
F}_k^{\nu}$ (\ref{eq_mot_sol}) of point $k$ of object $\nu$ is
then given by
\begin{align} \nonumber
{\bm F}_k^{\nu} &
 = \sum_{l=1}^{N_m} {\bm F}_{kl}^{\nu}({\bm r}_k^{\nu} -{\bm r}_l^{\nu}) \\
  \label{eq_motion} & +
\sum_{\mu=1}^{N_p}\sum_{l=1}^{N_m} \sum_{{\bm n}} {\bm F}_{kl}^{\nu \mu}({\bm r}_k^{\nu} -{\bm r}_l^{\mu} -{\bm R}_{{\bm n}}).
\end{align}
In general, infinite contributions of the intermolecular
interactions have be taken into account. For short-range
interactions, however, only nearest images contribute
significantly to the dynamics of a particle and we introduce a
potential cut-off, which is chosen such that self-interactions of
an object $\nu$ are prevented.\cite{alle:87}

\subsubsection{MPC fluid}

The stress tensor of the bare MPC solvent is obtained from Eq.
(\ref{eq_motion_vir_avr}). Evidently, either  fluid particles
themselves or their images are in the primary box. Denoting the
position (image or real) of a particle in the primary box by ${\bm
r}_i'(t)$, the particle position itself is given by ${\bm r}_i(t)
= {\bm r}_i'(t) + {\bm R}_i(t)$, where ${\bm R}_i(t)$ is the
lattice vector at time $t$. The force exerted on the particle
during the MPC collisions at times $t_q$ is
\begin{align} \label{force_time}
{\bm F}_i(t) = \sum_{q=0}^{\infty}
\Delta {\bm p}_i(t) \delta(t-t_q).
\end{align}
The time average of Eq. (\ref{eq_motion_vir_avr}) then reads
\begin{align} \nonumber
& \lla  F_{i \alpha} r_{i \beta} \rra  = \lim_{T \to \infty} \frac{1}{T}
\int_0^T F_{i \alpha}(t) r_{i \beta}(t) \ dt \\ \nonumber
& = \lim_{N \to \infty } \frac{1}{N}
\sum_{q=1}^{N}
\frac{1}{h}\int_{t_q-h}^{t_q} \Delta p_{i \alpha}(t)  r_{i \beta}(t) \delta (t - t_q) \ dt
\\ \nonumber
& = \lim_{N \to \infty } \frac{1}{N h} \sum_{q=1}^{N} \Delta p_{i \alpha}(t_q) r_{i \beta}(t_q)
\\ \nonumber
& = \lim_{N \to \infty } \frac{1}{N h} \sum_{q=1}^{N}[ \Delta p_{i \alpha}(t_q) r_{i \beta}'(t_q)
+ \Delta p_{i \alpha}(t_q) R_{i \beta}(t_q)]
\\ \label{instant_force}
& = \frac{1}{h} \lla \Delta p_{i \alpha}(t_q) r_{i \beta}'(t_q)
+ \Delta p_{i \alpha}(t_q) R_{i \beta}(t_q) \rra_{T} ,
\end{align}
where we introduced the average over collision steps
\begin{align} \label{def_average}
\lla \ldots \rra_{T} =  \lim_{N\to \infty} \lla \ldots \rra_{N} = \lim_{N\to \infty} \frac{1}{N} \sum_{q=1}^{N}
\ldots \ .
\end{align}
We define now an instantaneous external stress tensor
$\sigma_{\alpha \beta}^e$ by \cite{wink:92,wink:93}
\begin{align} \label{stress_external_mpc}
\sigma_{\alpha \beta}^e = \frac{1}{V h} \sum_{i=1}^{N_s} \Delta p_{i \alpha} R_{i \beta}.
\end{align}
Similarly, we introduce an instantaneous internal stress tensor by
\cite{wink:92,wink:93}
\begin{align} \label{stress_internal_mpc}
\sigma_{\alpha \beta}^i = -\frac{1}{V} \sum_{i=1}^{N_s} m \hat v_{i \alpha} \hat v_{i \beta}
-\frac{1}{Vh} \sum_{i=1}^{N_s} \Delta p_{i \alpha} r_{i \beta}'.
\end{align}
According to Eqs. (\ref{eq_motion_vir_avr}) and
(\ref{instant_force}), the averages of the two terms are equal,
i.e., $\langle \sigma_{\alpha \beta}^e \rangle_T = \langle
\sigma_{\alpha \beta}^i \rangle_T$.  Hence, we obtain two
equivalent expressions for the stress tensor. Equation
(\ref{stress_external_mpc}) corresponds to the mechanical
definition as force per area ($R_{i \beta} \sim L$) and Eq.
(\ref{stress_internal_mpc}) follows from the momentum flux across
a surface. Correspondingly, the external stress tensor includes
only force terms, i.e., collisional contributions, whereas the
internal stress tensor comprises kinetic and collisional
contributions.

In general, the pressure follows from the stress tensor via the
relation $p= -\sum_{\alpha} \sigma_{\alpha \alpha} /3$.

\begin{figure}[t]
\includegraphics*[width=7.5cm,clip]{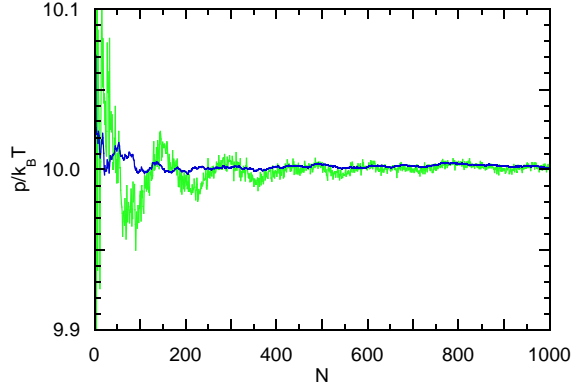}
  \caption{Internal $\lla p^i \rra_N$ (blue) and external $\lla p^e \rra_N$
  (green, large fluctuations) pressure as
  a function of the
  number of collision steps $N$. The collision time is $\tilde h = 0.1$ and the
  time $t = N h$.}
  \label{fig_press_per}
\end{figure}

Figure \ref{fig_press_per} displays the dependencies of the
averages $\lla p^i \rra_N$, $\lla p^e \rra_N$ (cf. Eq.
(\ref{def_average})) of the internal and external pressures on the
number of collision steps, which yield  the macroscopic pressure
$p = \lla p^i \rra_T = \lla p^e \rra_T$ in  the limit $N \to
\infty$. Evidently, both expressions approach the same limiting
value for a large number of collision steps. The fluctuations of
the average external pressure are larger, since the number of
particles included in the pressure calculation are smaller as
compared to the internal pressure. Moreover, the fluctuations of
$p^e$ itself are larger and increase like the square root of $t$
with time, because the fluid particles diffuse through the
infinite periodic system.\cite{wink:92} The pressure is given by
the kinetic contribution $p = \sum_{i=1}^{N_s} m \lla \hat {\bm
v}_i \rra_T/(3V) = N_c k_BT$. This follows from the fact that the
momentum change in a collision cell is independent of the actual
positions of the particles, hence $\Delta {\bm p}_i$ and ${\bm
r}_i$ are uncorrelated and the collisional contributions to the
internal pressure/stress vanish.

\subsubsection{MPC fluid and embedded particles}

For the case of embedded particles, Eq. (\ref{eq_motion_vir_avr})
reads
\begin{align} \nonumber
0 & =  \sum_{i=1}^{N_s} m \lla  v_{i \alpha} v_{i \beta} \rra_{T} +
 \sum_{\nu=1}^{N_p} \sum_{k=1}^{N_m} M_k \lla \hat v_{k \alpha}^{\nu}
 \hat v_{k \beta}^{\nu} \rra_{T}
\\ \nonumber & +
\frac{1}{h}  \sum_{i=1}^{N_s}
\lla \Delta p_{i\alpha} r_{i\beta} \rra_{T}  +
\frac{1}{h}  \sum_{\nu=1}^{N_p} \sum_{k=1}^{N_m}
\lla \Delta p_{k\alpha}^{\nu} r_{k\beta}^{\nu} \rra_{T} \\ \nonumber &
+ \frac{1}{2} \sum_{\nu=1}^{N_p} \sum_{k,l=1}^{N_m} \lla F_{kl \alpha}^{\nu}
[r_{k \beta}^{\nu} - r_{l \beta}^{\nu}] \rra_{T} \\ \label{eq_vir_emb_avr}&
+ \frac{1}{2} \sum_{\nu,\mu=1}^{N_p} \sum_{k,l=1}^{N_m} \sum_{{\bm n}} \lla F_{kl \alpha}^{{\bm n}\nu\mu}
[r_{k \beta}^{\nu} - r_{l \beta}^{\mu}] \rra_{T},
\end{align}
when we use Eqs.  (\ref{eq_motion}) and (\ref{instant_force}), and
the abbreviation $F_{kl \alpha}^{{\bm n}\nu\mu} = F_{kl
\alpha}^{\nu\mu}({\bm r}_k-{\bm r}_{l} - {\bm R}_{\bm n})$.

As described in Refs.
\onlinecite{wink:92,wink:93,wink:02,theo:93,akke:04} for the
solute and with the same strategy as for the bare solvent, we
obtain the following instantaneous stress tensors
\begin{align} \nonumber
\sigma_{\alpha \beta}^{e}  & =  \frac{1}{2V} \sum_{\nu, \mu =1}^{N_p} \sum_{k,l=1}^{N_m}
\sum_{{\bm n}} F_{kl\alpha}^{{\bm n}\nu\mu} R_{{\bm n} \beta} \\ \label{stress_embed_ex} & +
\frac{1}{Vh}  \sum_{i=1}^{N_s}
\Delta p_{i \alpha} R_{i \beta}+
\frac{1}{Vh} \sum_{\nu=1}^{N_p} \sum_{k=1}^{N_m}
\Delta p_{k\alpha}^{\nu} R_{k\beta}^{\nu},
\\ \nonumber
\sigma_{\alpha \beta}^{i}  & = - \frac{1}{V} \sum_{i=1}^{N_s}  m \hat v_{i \alpha} \hat v_{i \beta}
- \frac{1}{V} \sum_{\nu=1}^{N_p} \sum_{k=1}^{N_m} M_k \hat v_{k \alpha}^{\nu} \hat v_{k \beta}^{\nu}
\\ \nonumber & -
\frac{1}{Vh} \sum_{i=1}^{N_s}
\Delta p_{i \alpha} r_{i \beta}' -
\frac{1}{Vh} \sum_{\nu=1}^{N_p} \sum_{k=1}^{N_m}
\Delta p_{k\alpha}^{\nu} {r'}_{k\beta}^{\nu}
\\ \nonumber & -
\frac{1}{2V} \sum_{\nu=1}^{N_p} \sum_{k,l=1}^{N_m}  F_{kl \alpha}^{\nu}
[r_{k \beta}^{\nu} - r_{l \beta}^{\nu}] \\ \label{stress_embed_int}&
- \frac{1}{2V} \sum_{\nu,\mu=1}^{N_p} \sum_{k,l=1}^{N_m} \sum_{{\bm n}}  F_{kl \alpha}^{{\bm n}\nu\mu}
[r_{k \beta}^{\nu} - r_{l \beta}^{\mu} -R_{{\bm n} \beta}] .
\end{align}
Again, the averages are equal $\langle \sigma_{\alpha \beta}^e
\rangle_T =\langle \sigma_{\alpha \beta}^i \rangle_T$. The
solvent-solute coupling is captured in the terms with the momenta
$\Delta p_{k \alpha}^{\nu}$ of the monomers as well as in those
for the fluid momenta $\Delta p_{i \alpha}$ of collision cells
containing monomers.

\subsection{Confining walls}

We will now determine the stress tensors for an MPC fluid confined
between two solid walls. The walls are parallel to the $xy$-plane
and periodic boundary conditions are applied along the $x$- and
$y$-directions. The center of the reference system is located in
the middle between the two walls, i.e., the wall positions are
$z_w = \pm L/2$. The equations of motion of the fluid particles
are then modified by the wall interactions. We will assume no-slip
boundary conditions, which we realize by the bounce-back rule,
i.e., the velocity of a fluid particle is reverted when it hits a
wall (${\bm v}_i \to - {\bm v}_i$).\cite{lamu:01}

\begin{figure}[t]
\includegraphics*[width=7.5cm,clip]{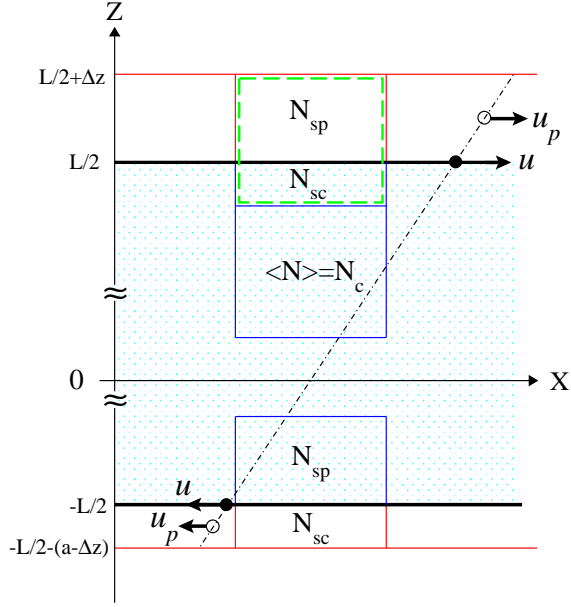}
  \caption{Illustration of the random shift and the distribution of
  particles in collision cells cut by the walls. The walls are located
  at $z=\pm L/2$. Under shear, the walls move with the velocities
  $u = \pm \dot \gamma L/2$. The phantom particles are located in the
  centers of the truncated parts of the collision cell at $(L+\Delta z)/2$ and
  $-(L+a-\Delta z)/2$, respectively. They move with the velocities
  $u_p = \dot \gamma (L+\Delta z)/2$ and
  $u_p = -\dot \gamma (L+a -\Delta z)/2$, respectively. The dashed-dotted
  line indicates the linear velocity profile.}
  \label{fig_shift}
\end{figure}

The random shift perpendicular to the walls is implemented as
follows. Without random shift, the most upper and lower border of
the collision cells coincides with the respective wall. To enable
a random shift, an additional layer of (empty) collision cells is
added below the lower wall. In a random shift, the whole collision
lattice is shifted in the positive $z$-direction by a uniformly
distributed displacement $\Delta z$, with $0 \le \Delta z \le a$,
as illustrated in Fig. \ref{fig_shift}. The random shift typically
leads to partially occupied cells at the walls, which in turn
causes a violation of the no-slip boundary condition under shear
\cite{lamu:01}. To restore no-slip boundary conditions, typically
virtual particles are added to every cell cut by a wall and
occupied by a number of particles $N_{sc}$ smaller than the
average number of particles $N_c$, such that the average particle
density is restored. However, this does not completely prevent
slip, because the average center-of-mass position of all particles
in a collision cell---including the phantom particle---does not
coincide with the wall. In order to fully account for the no-slip
boundary condition, we propose the following modification of the
original approach. To treat a surface cell on the same basis as a
cell in the bulk, i.e., the number of particles satisfies a
Poisson distribution with the average $N_c$, we take fluctuations
in the particle number into account by adding $N_{sp}$ particles
to every cell cut by a wall such that $\lla N_{sp}+N_{sc} \rra =
N_c$. The momentum ${\bm P}$ of a virtual particle is taken from
the Maxwell-Boltzmann distribution with the variance $\sigma^2 = m
N_{sp} k_B T$ and, at equilibrium, zero average. (The case of a
shear flow is discussed in Sec. V B.) There are various ways to
determine the number $N_{sp}$. For a system with two parallel
walls, we suggest to use the number of fluid particles in the
surface cell cut by the opposite wall. The average of the two
numbers is equal to $N_c$. Alternatively, $N_{sp}$ can be taken
from a Poisson distribution with average $N_c$ accounting for the
fact that there are already $N_{sc}$ particles in the cell.
Collisions are then performed with all the particles in the cells.
The center-of-mass velocity of the particles in a boundary cell is
\begin{align} \label{cm_phantom}
{\bm v}_{cm} = \frac{1}{ m (N_{sc} +N_{sp})} \left( \sum_{i=1}^{N_{sc}}m \hat{\bm v}_i +
{\bm P}\right) .
\end{align}
Naturally, this type of collisions will affect the external stress
tensor.

The total force on a fluid particle $i$ comprises contributions
from collisions among particles and collisions with the walls,
i.e.,
\begin{align} \label{force_surface}
{\bm F}_i = \sum_{q=0}^{\infty} \Delta {\bm p}_i \delta(t - t_q) +
\sum_{q=0}^{\infty} \Delta {\bm p}_i^w \delta(t - t_q^w),
\end{align}
where $t_q^w$ is the time at which the particle hits a wall and
$\Delta {\bm p}_i^w = -2 m {\bm v}_i$ its momentum change. By
averaging over a collision interval, the force term in Eq.
(\ref{eq_motion_vir_avr}) becomes
\begin{align} \nonumber
\lla \sum_{i=1}^{N_s} F_{i \alpha} r_{i \beta} \rra_{h} & = \frac{1}{h} \sum_{i=1}^{N_s}
\Delta p_{i \alpha} r_{i \beta}
\\ \label{virial_wall} &
+ \frac{L}{h} \sum_{i=1}^{N_s}
\Delta p_{i \alpha}^w \left[\Theta(r_{i z})-\frac{1}{2} \right] \delta_{z \beta},
\end{align}
with the Heaviside function
\begin{align} \label{heaviside}
\Theta(x) = \left\{
\begin{array}{cc}
1, & x > 0 \\
0, & x < 0
\end{array}
. \right.
\end{align}
By adding and subtracting the term $\sum_{i \in bc} \Delta p_{i
\alpha} [\Theta(r_{i z})-\frac{1}{2} ]  L \delta_{z \beta}$ in Eq.
(\ref{virial_wall}), we obtain the instantaneous external and
internal stress tensors
\begin{align} \nonumber
\sigma_{\alpha \beta}^e  = &
 \frac{1}{V h} \sum_{i=1}^{N_s} \Delta p_{i \alpha} R_{i \beta} [1-\delta_{z \beta}] \\ \nonumber
& + \frac{L}{V h} \sum_{i=1}^{N_s}
\Delta p_{i \alpha}^w \left[\Theta(r_{i z})-\frac{1}{2} \right] \delta_{z \beta} \\
 \label{stress_external_wall_mpc}
& + \frac{L}{V h} \sum_{i \in bc} \Delta p_{i \alpha} \left[ \Theta(r_{i
z})-\frac{1}{2} \right]   \delta_{z \beta} , \\ \nonumber
\sigma_{\alpha \beta}^i  = & -\frac{1}{V} \sum_{i=1}^{N_s} m \hat v_{i \alpha} \hat v_{i \beta}
-\frac{1}{Vh} \sum_{i =1}^{N_s} \Delta p_{i \alpha} r_{i \beta}'
\\ \label{stress_internal_wall_mpc}
& + \frac{L}{V h} \sum_{i \in bc} \Delta p_{i \alpha} \left[ \Theta(r_{i z})-\frac{1}{2} \right]
\delta_{z \beta},
\end{align}
with $r_{iz}' = r_{iz}$, including both, the contributions by the
surfaces as well as by the periodic boundary conditions. The
surface contribution to the stress tensor in Eq.
(\ref{stress_external_wall_mpc}) clearly shows that stress is
force/area. \cite{wink:93,tao:08} The last term of Eq.
(\ref{stress_internal_wall_mpc}) originates from the finite range
of the fluid-surface interaction. In MPC, the non-locality of the
fluid collisions is responsible for the extended range. The
collision interaction of a particle with the surface is of zero
range, whereas for a finite range potential, e.g., a Lennard-Jones
potential, another term would appear as discussed in Ref.
\onlinecite{wink:93}.

Simulations yield a similar time dependence of the pressure as for
the periodic system, which is display in Fig. \ref{fig_press_per}.

We will not explicitly discuss the inclusion of a solute in the
calculation of the stress tensor here. A detailed derivation of
the stress tensors for mixed confined and periodic molecular
systems is presented in Ref. \onlinecite{wink:93} and the
contribution of the solvent-solute interaction is identical to the
terms presented in Eqs. (\ref{stress_embed_ex}) and
(\ref{stress_embed_int}).

\section{Stress tensor: Shear Flow}

The presence of shear flow alters some of the terms of the fluid
stress tensors.  Therefore, we will discuss this type of external
field in more detail.

In general, shear is applied in the $x$-direction and the gradient
direction is along the $z$-axis.

\subsection{Periodic boundary conditions}

Again, we will discuss a system with periodic boundary conditions
first. Because of the external field, the time average of the left
hand side of Eq. (\ref{eq_motion_vir}) does not vanish anymore.
Neglecting fluctuations for the moment, the velocity $v_{ix} $ of
the linear flow profile is $v_{ix} = \dot \gamma r_{iz}$, with the
shear rate $\dot \gamma$. The time average $\lla d (v_{ix} r_{iz})
dt \rra$ is then given by $\lim_{T \to \infty} r_{iz}^2(T)/T $.
Since a particle is diffusing along the gradient direction,
$r_{iz}^2(T) \sim T$ and the average is finite. In order to arrive
at a vanishing term, we subtract the derivative of the velocity
profile $d(\dot \gamma r_{i z})/dt = \dot \gamma v_{iz}$ from both
sides of Eq. (\ref{eq_motion_com}). This leads to the modified
equation
\begin{align} \nonumber
\frac{d}{dt} \sum_{i=1}^{N_s} m (v_{i x}-\dot \gamma r_{i z})  r_{i z} = &
\sum_{i=1}^{N_s} m (v_{i x}-\dot \gamma r_{iz}) v_{i z}  \\ \label{eq_motion_vir_shear} & +
\sum_{i=1}^{N_s} F_{i x} r_{i z} - \dot \gamma \sum_{i=1}^{N_s} m v_{iz} r_{iz} .
\end{align}
Evidently, the (time) average of the left hand side vanishes.
Applying the definition of the time average (\ref{instant_force}),
the velocity  terms on the right hand side read as
\begin{align} \nonumber
 \lla (v_{i x}-\dot \gamma r_{i z})  v_{i z} \rra & =
\lla  \hat v_{iz} \hat v_{ix}' \rra_{T} + \frac{\dot \gamma h }{2} \lla  \hat v_{iz}^2 \rra_{T} ,
\\ \lla v_{i z}  r_{i z} \rra & = \frac{1}{2}  \lla
(v_{iz}+\hat v_{iz}) r_{iz}  \rra_{T}
\end{align}
in the stationary state. Note that $\hat {\bm v}_{i}(t_q)$ is the
velocity before the collision and ${\bm v}_{i} (t_q)$ that after
the collision. Similar to the notation for the positions,
$v_{ix}'$ denotes the velocity in the primary cell of the periodic
system, i.e., $v_{ix} = v_{ix}' + \dot \gamma R_{ix}$. (The
particle velocities along the other spacial directions are
identical for each periodic image.) The original expression $\lla
(\hat v_{i x}-\dot \gamma r_{i z}) \hat v_{i z} \rra_{T}$ reduces
to $\lla \hat v_{i x}' \hat v_{i z} \rra_{T}$, because the average
$\lla \hat v_{iz} r_{iz}'\rra_{T}$ vanishes. We like to point out
that the change from ${\bm v}_i$, ${\bm r}_i$ to ${\bm v}_i'$,
${\bm r}_{i}'$ corresponds to the application of Lees-Edwards
periodic boundary conditions in non-equilibrium simulations of
simple shear. For the sake of completeness, we emphasize that the
time and ensemble average of the last term on the right hand side
of Eq. (\ref{eq_motion_vir_shear}) is $\sum_{i=1}^{N_s} m \dot
\gamma \lla  v_{iz} r_{iz} \rra = m \dot \gamma  N_s D$, where $D$
is the diffusion coefficient of an MPC particle.

We are now in the position to define instantaneous external and
internal stress tensors as
\begin{align}  \label{stress_ex_shear}
\sigma_{xz}^{e}  & =
\frac{1}{Vh}  \sum_{i=1}^{N_s}
\Delta p_{i x} R_{i z}  - \frac{\dot \gamma}{2V} \sum_{i=1}^{N_s}  m
(v_{iz}+\hat v_{iz}) R_{iz} ,
\\ \label{stress_in_shear}
\sigma_{xz}^{i}  & = - \frac{1}{V} \sum_{i=1}^{N_s}  m \hat v_{i x}' \hat v_{i z}
 - \frac{\dot \gamma h}{2V} \sum_{i=1}^{N_s} m v_{iz}^2
 -\frac{1}{Vh}  \sum_{i=1}^{N_s} \Delta p_{i x} r_{iz}',
\end{align}
which obey the relation $\lla \sigma_{xz}^i \rra_T = \lla
\sigma_{xz}^e \rra_T$. The presence of the external field leads to
additional terms contribution to the stress tensors compared to
the expressions (\ref{stress_external_mpc}) and
(\ref{stress_internal_mpc}). The extra term in $\sigma_{xz}^i$
results from the streaming dynamics and vanish in the limit $h \to
0$. Since a discrete time dynamics is fundamental for the MPC
method, the collision time will always be finite.

\begin{figure}[t]
\includegraphics*[width=7.5cm,clip]{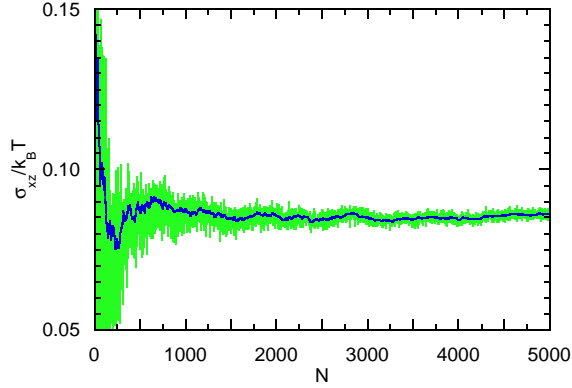}
  \caption{Internal $\lla \sigma_{xz}^i \rra_N$ (blue) and external
  $\lla \sigma_{xz}^e \rra_N$ (green, large fluctuations) stress tensor
  as function of the
  number of collision steps. The collision time is $\tilde h = 0.1$.
  At $t=0$, the system is in a stationary state.}
  \label{fig_stress_per}
\end{figure}

An example of the time dependence of the internal  and external
stress tensors, i.e., $\lla \sigma^i_{xz} \rra_N$, $\lla
\sigma^e_{xz} \rra_N$, under shear is shown in Fig.
\ref{fig_stress_per}. Both expressions approach the same limiting
value for a large number of collision steps. The fluctuations of
the external stress tensor component are again larger.

\subsection{Confining walls}

For the system described in Sec. III B, shear is imposed by the
opposite movement of the confining walls with the velocities $u =
\dot \gamma z_w = \pm \dot \gamma L/2$. The structure of the
external stress tensor (\ref{stress_external_wall_mpc}) is
maintained. However, the wall momentum $\Delta p_{ix}^{w}$ changes
to $\Delta p_{ix}^{w}= -2 m v_{iz} + 2m u$. In addition, the
phantom particles of the partially filled surface cells possess a
finite velocity. Hence, the momentum $P_x$ is now determined from
the Maxwell-Boltzmann distribution of the same variance as before,
but with the velocity
\begin{align} \label{mean_momentum}
u_p =m N_{sp} \left(u+  \frac{\dot \gamma}{2} \Delta z- \frac{\dot \gamma}{2} a
[1-\Theta(z_w)] \right).
\end{align}
As illustrated in Fig. \ref{fig_shift}, $\Delta z$ is the fraction
of the cells truncated by the wall at $z_w=L/2$. Correspondingly,
$a-\Delta z$ is the fraction of the cells truncated by the
opposite wall. In our description, the phantom particles are
located in the centers (along the $z$-axis) of the truncated parts
of a surface cells. The advantage of this approach over the
previous implementation is that the center-of-mass velocity of a
surface cell is equal to the velocity of the wall, as shown in
Appendix A. This implies no-slip at the wall.

For the internal stress tensor, the time average of the terms
$\hat v_{ix} \hat v_{iz}$ over a collision interval is modified.
The integral now becomes
\begin{align}  \label{av_wall_kin}
\frac{1}{h} \int_{t_q-h}^{t_q} v_{ix}(t) v_{iz}(t) dt  =
\hat v_{ix}(t_q) \hat v_{iz}(t_q) - \frac{2 u}{h} \hat v_{iz}(t_q) \Delta t_q^i
\end{align}
for a particle which collides with a wall at $t_q^w$ in the
interval $t_q -h < t_q^w < t_q$ and $\Delta t_q^i/h =
1-(t_q-t_q^w)/h$. Evidently, the average over $v_{iz}$ is
non-zero, because the relevant particles move always towards the
respective surface. Hence, $\sigma_{xz}^i$ becomes
\begin{align} \nonumber
\sigma_{xz}^i  = & -\frac{1}{V} \sum_{i=1}^{N_s} m \hat v_{i x} \hat v_{i z}
+ \frac{2u}{Vh} \sum_{i=1}^{N_s} m \hat v_{iz} \Delta t_q^i \\ \nonumber
& -\frac{1}{Vh} \sum_{i=1}^{N_s} \Delta p_{i x} r_{i z}
\\ \label{stress_internal_wall_mpc_shear}
& + \frac{L}{V h} \sum_{i \in bc}
\Delta p_{i x}  \left[ \Theta(r_{i z})-\frac{1}{2} \right]   \ .
\end{align}

The other components of the stress tensor are obtained via  Eq.
(\ref{stress_internal_wall_mpc}).

Simulations confirm that the center of mass velocity of the
particles interacting with phantom particles in the surface cells
are indeed equal to the velocity of the respective surface.
Moreover, the time dependent averages of the internal and external
stress tensors $\lla \sigma^i_{xz} \rra_N$, and $\lla
\sigma^e_{xz} \rra_N$ are similar to those displayed in Fig.
\ref{fig_stress_per}.

\section{Viscosity}

The derived expressions for the stress tensors are independent of
any particular collision rule. Transport coefficients such as the
viscosity of a system, however, depend on the apply collision
procedure.

Analytical expressions for the viscosity of an MPC fluid have been
derived by various approaches.
\cite{ihle:01,male:00,kiku:03,ihle:05,nogu:08,kapr:08,gomp:08_p}
Since the stress tensors of Eqs. (\ref{stress_ex_shear}),
(\ref{stress_in_shear}), and
(\ref{stress_internal_wall_mpc_shear}) are novel, we will here
derive the viscosity based on these expressions for the stochastic
rotation version of MPC described in Sec. II.

In simple shear flow with the velocity field $ v_x = \dot \gamma
z$, the viscosity $\eta$ is related to the stress tensor via $\eta
= \sigma_{xz}/\dot\gamma$, where the (macroscopic) stress tensor
follows from $\sigma_{xz} = \lla \sigma_{xz}^i \rra_{T} = \lla
\sigma_{xz}^e \rra_{T}$. For an MPC fluid, the stress tensor is
composed of a kinetic and collisional contribution,
\cite{ihle:01,male:00,kiku:03,nogu:08,kapr:08,gomp:08_p} i.e,
$\sigma_{xz} =
\sigma_{xz}^{\mathrm{kin}}+\sigma_{xz}^{\mathrm{col}}$, which
implies that the viscosity $\eta = \eta_{\mathrm{kin}}+
\eta_{\mathrm{col}}$ consists of a kinetic $\eta_{\mathrm{kin}}$
and collisional $\eta_{\mathrm{col}}$ part too.
\cite{ihle:01,male:00,kiku:03,nogu:08,kapr:08,gomp:08_p}

\subsection{Periodic boundary conditions}

For a system with periodic boundary conditions, the two
contributions to the viscosity are conveniently obtained from the
internal stress tensor (\ref{stress_in_shear}).

The kinetic contribution $\eta_{\mathrm{kin}}$ is determined by
the streaming step, i.e., velocity dependent terms in Eq.
(\ref{stress_in_shear}). To find the mean $\lla \hat v_{ix}' \hat
v_{iz}' \rra$, we consider a complete MPC dynamics step. The
velocity $v_{ix}'(t_q)$ before streaming is related to the
velocity $\hat v_{ix}'(t_q+h)$ after streaming via $\hat
v_{ix}'(t_q+h) = \hat v_{ix}(t_q+h) - \dot \gamma r_{iz}(t_q+h) =
v_{ix}(t_q) - \dot \gamma r_{iz}(t_q) - \dot \gamma v_{iz}(t_q)h =
v_{ix}'(t_q) - \dot \gamma v_{iz}(t_q)h $. With $\hat
v_{iz}'(t_q+h) = v_{iz}(t_q)$, we obtain the average
\begin{align} \label{rel_stream}
\lla \hat v_{ix}'(t_q+h) \hat v_{iz}(t_q+h) \rra    =
\lla v_{ix}'(t_q)  v_{iz}(t_q) \rra - \dot \gamma h \lla v_{iz}^2 \rra.
\end{align}
Here, the average comprises both, a time average and an ensemble
average over the orientation of the rotation axis. The velocities
after streaming are changed by the subsequent collisions, which
yields,  with the corresponding momenta of the rotation operator
${\mathbf R}(\alpha)$,  $\lla v_{ix}'(t) v_{iz}(t) \rra = f \lla
\hat v_{ix}'(t) \hat v_{iz}(t)\rra $ and $f=1+(1-1/N_c)(2 \cos(2
\alpha)+ 2 \cos (\alpha) - 4)/ 5$. \cite{kiku:03,nogu:08} Note,
velocity correlations between different particles are neglected,
i.e., molecular chaos is assumed. Thus, in the steady stead [$\lla
\hat v_{ix}'(t) \hat v_{iz}'(t)\rra = \lla \hat v_{ix}'(t+h) \hat
v_{iz}'(t+h) \rra$], we find
\begin{align} \label{vel_corr}
\lla \hat v_{ix}' \hat v_{iz} \rra_{N} = - \frac{\dot \gamma h }{1-f} \lla v_{iz}^2 \rra
\end{align}
by using Eq. (\ref{rel_stream}). Hence, with the equipartition of
energy $\lla v_{iz}^2 \rra = k_BT/m$, the kinetic viscosity is
given by
\begin{align} \label{visc_kin}
\eta_{\mathrm{kin}} = \frac{N_s k_B T h}{V}  \left[ \frac{5 N_c}{(N_c-1)(4 - 2 \cos(\alpha) - 2 \cos (2\alpha)  )} - \frac{1}{2} \right],
\end{align}
in agreement with previous calculations.

The collisional viscosity is determine by the momentum change of
the particles during the collision step. Since the collisions in
the various cells are independent, it is sufficient to consider
one cell only. The positions of the particles of that cell can be
expressed as ${\bm r}_i' = {\bm r}_c + \Delta {\bm r}_i$, where
${\bm r}_c$ is chosen as the center of the cell. Because of
momentum conservation, the term $\sum_{i=1}^{N_c} \Delta p_{ix}
r_{iz}'$ then reads as $\sum_{i=1}^{N_c} \Delta p_{ix} \Delta
r_{iz}$. The averages over thermal fluctuations and random
orientations of the rotation axis yield
\begin{align} \nonumber
& \lla \Delta p_{ix} \Delta r_{iz} \rra = \frac{2m \dot \gamma}{3} (\cos (\alpha)-1) \\
\label{eta_col_aux}
&\times \left[ \left( 1 - \frac{1}{N_c} \right) \lla \Delta r_{iz}^2 \rra - \frac{1}{N_c}
\sum_{j \ne i =1}^{N_c} \lla \Delta r_{iz} \Delta r_{jz} \rra \right].
\end{align}
The average over the uniform distribution of the positions within
an cell yields $\lla \Delta r_{iz} \Delta r_{jz} \rra =0$ for $i
\ne j$ and
\begin{align} \label{pos_av}
\frac{1}{a} \int_{-a/2}^{a/2} \Delta r_{iz}^2 dz = \frac{a^2}{12}.
\end{align}
Hence, the collisional viscosity is given by
\begin{align} \label{visc_col}
\eta_{\mathrm{col}} = \frac{N_s m a^2}{18 V h}  \left ( 1 -  \cos(\alpha) \right)
\left(1 - \frac{1}{N_c} \right) ,
\end{align}
again in agreement with previous calculations.

Here, we assume that the number of particles in a collision cell
$N_c$ is sufficiently large ($N_c > 3$) to neglect fluctuations.
\cite{gomp:08_p} For a small number of particles, density
fluctuations have to be taken into account. Then, Eqs.
(\ref{visc_kin}) and (\ref{visc_col}) have to be averaged over the
particle number using a Poisson distribution with the mean value
$N_s/V$.\cite{kiku:03,gomp:08_p}

We perform simulations for various MPC parameters and found a very
good agreement between the viscosities determined via Eqs.
(\ref{stress_ex_shear}), (\ref{stress_in_shear}) and the
analytical expression Eqs. (\ref{visc_kin}) and (\ref{visc_col}).

\subsection{Confining walls}

Under confinement, the component $\sigma_{xz}^e$ of the external
stress tensor is determined by the collisions of the fluid
particles with the walls, which corresponds to the kinetic
contribution, and the collisions of fluid particles within the
partially filled surface cells, which yields the collisional
contribution to the viscosity. Since the averages over the stress
tensor contributions from each wall are equal, we find
\begin{align} \nonumber
\lla \sigma_{xz}^e \rra = & \sigma_{xz}^{\mathrm{kin}} + \sigma_{xz}^{\mathrm{col}}
\\ \label{stress_split}
= & \frac{L}{Vh} \lla \sum_{i=1}^{N_s}  \Delta p_{ix}^w \rra
+ \frac{L}{Vh} \lla \sum_{i \in bc}  \Delta p_{ix} \rra,
\end{align}
and the averages are taken over one surface only.

As shown in Appendix B and C, the evaluation of the averages
yields exactly the same expressions for the viscosities  as
derived for a periodic system in Sec. IV A, namely Eqs.
(\ref{visc_kin}) and (\ref{visc_col}).  We like to point out that
this is not true in general. A simulation study with the mean of
the momentum $P_x = m (N_c - N_{sc})u$ yields different results
for the two viscosity contributions, although the total viscosity
agrees with the theoretical prediction. \cite{tao:08} Only for our
choice of the mean momentum (\ref{mean_momentum}) follows
agreement with the theoretical expressions.

\begin{figure}[t]
\includegraphics*[width=7.5cm,clip]{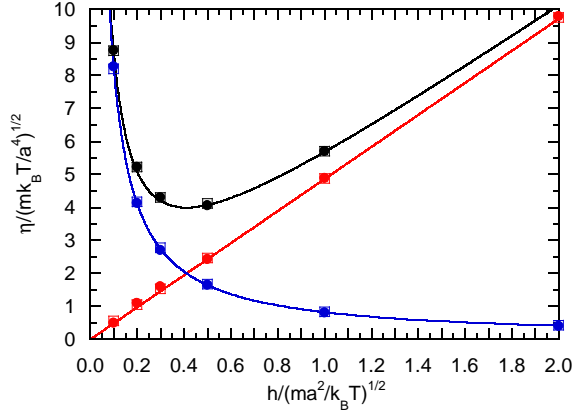}
  \caption{Viscosities determined  via the internal (bullets) and
  external (open squares) stress tensors for a system confined between
  walls as function of the collision time. The analytical results for
  the total (black), the kinetic (red, $\sim h$),
  and collisional (blue, $\sim 1/h$)
  contributions are presented by solid lines.}
  \label{fig_viscosity}
\end{figure}

Figure \ref{fig_viscosity} depicts viscosities determined via the
internal $\lla \sigma_{xz}^i \rra_T$
(\ref{stress_internal_wall_mpc_shear}) and external $\lla
\sigma_{xz}^e \rra_T$ (\ref{stress_split}) stress tensors and
their respective collisional and kinetic contributions. Evidently,
the averages agree very well with each other. Moreover, the
simulation results agree very well with the analytical predictions
for the kinetic and collisional viscosities of Eqs.
(\ref{visc_kin}) and (\ref{visc_col}).

\section{Summary}

In this article, we have been introducing external and internal
stress tensors for an MPC fluid resembling those of atomistic
molecular fluids. Systems with periodic boundary conditions and
fluids confined in a slit have been addressed and their
peculiarities have been worked out. Moreover, the modifications of
the stress tensors caused by the presence of simple shear have
been determined. Based on the derived stress tensors, an
analytical expressions for the viscosity has been derived, which
agrees with previous
results.\cite{ihle:01,male:00,kiku:03,ihle:05,nogu:08,kapr:08,gomp:08_p}
In addition, stress tensors for systems containing solute
molecules are presented, which are coupled to the solvent in the
MPC collision step. These expressions explicitly comprise the
solvent-solute contributions to the stress tensors.

The stress tensors can easily be modified to account for a
different coupling between the solvent and the solute. In Refs.
\onlinecite{male:00,lee:04,kapr:08}, the solute interacts through
an intermolecular potential, i.e., the Lennard-Jones potential,
with the solvent. This results in an additional virial term in the
stress tensor---similar to the solute intermolecular
interactions---with the forces between the solvent and the solute
particles and their respective positions.

Simulations for various MPC parameters confirm the equivalence of
time averages of the internal and external stress tensors of the
fluid for both types of boundary conditions. Moreover, the
calculated viscosities are in accord with the corresponding
analytical expressions.

The stress tensors can easily be calculated, since they require
known quantities, i.e, positions, velocities, and momenta changes,
only. Moreover, all particles contribute in the calculation of the
internal stress tensors and no extra hypothetical plane needs to
be introduced.\cite{kiku:03,nogu:08}

\acknowledgments

C.-C. H. gratefully thanks J. P. Ryckaert and G. Desrt\'{e}e of U.
L. B., Belgium for valuable discussions and technical support.
Financial support by the German Research Foundation (DFG) within
SFB TR6 is gratefully acknowledged.

\appendix

\section{Center-of-mass velocity in surface cells}

The center-of-mass velocity of all particles in a surface cell
truncated by a wall and filled with a phantom particle of momentum
${\bm P}$ is given by Eq. (\ref{cm_phantom}). The average of the
component in the flow direction for the wall at $z_w=L/2$ reads
\begin{align} \label{app_cm_vel_1}
\lla {v}_{cm,x} \rra = \frac{1}{N_c} \lla  \sum_{i=1}^{N_{sc}} \hat v_{ix} +
 N_{sp} \left[u + \frac{\dot \gamma}{2} \Delta z \right] \rra ,
\end{align}
where $N_c= N_{sc}+N_{sp}$. The fluctuations of $P_{x}$ have been
averaged out already. The average over the fluctuations of the
fluid particle velocities yields
\begin{align} \nonumber
\lla {v}_{cm,x} \rra = & \frac{1}{N_c}  \lla N_{sc} \dot \gamma \bar z +
 N_{sp} \left[u + \frac{\dot \gamma}{2} \Delta z \right] \rra   \\ \nonumber
 = &  \frac{1}{N_c} \lla N_{sc} \left[u - \frac{\dot \gamma}{2} (a-\Delta z) \right]
 + N_{sp} \left[u + \frac{\dot \gamma}{2} \Delta z \right] \rra  \\ \label{app_cm_vel_2}
 = &  u +  \frac{\dot \gamma}{2} \lla \Delta z \rra   -
 \frac{\lla N_{sc} \rra}{2N_c}  \dot \gamma a .
\end{align}
$a-\Delta z$ is the part of the intersected surface cell which is
within the fluid slit, and the average $\bar z$ of the particle
position in a cell is
\begin{align} \label{app_cm_vel_3}
\bar z = \frac{1}{a-\Delta z} \int_{L/2 - a + \Delta z}^{L/2} z dz = \frac{L}{2} -
\frac{a-\Delta z}{2} .
\end{align}

 The remaining average is over the random
shift $\Delta z$ and the particle number $N_{sc}$. The average of
$\Delta z$, $0 \le \Delta z \le a$, yields $\lla \Delta z \rra =
a/2$ and $\lla N_{sc} \rra = N_c/2$. Thus, we find $\lla
v_{cm,x}\rra=u$.

\section{Surface collisional stress tensor}

The collisional contribution to the stress tensor
(\ref{stress_split}) at a wall is given by
\begin{align} \label{app_col_1}
\sigma_{xz}^{\mathrm{col}} =\frac{1}{a^2 h} \lla \sum_{i=1}^{N_{sc}} \Delta p_{ix} \rra,
\end{align}
because the contributions from the various cells are independent.
Averaging over the orientation of the rotation axis  and the
fluctuations of the momenta yields
\begin{align} \nonumber
\sigma_{xz}^{\mathrm{col}} = & \frac{2(\cos(\alpha)-1) m}{3 a^2 h} \lla
\sum_{i=1}^{N_{sc}} v_{ix} -N_{sc} v_{cm,x} \rra \\ \nonumber = &
\frac{2(\cos(\alpha)-1)}{3 a^2 h N_c} \lla N_{sp} \sum_{i=1}^{N_{sc}} m v_{ix} -
N_{sc} P_{x} \rra \\ \nonumber
= & \frac{2 (\cos(\alpha)-1) \dot \gamma m}{3 a^2 h N_c} \lla N_{sp} N_{sc}
\left[\bar z - \frac{1}{2} \left(L + \Delta z \right) \right] \rra \\ \label{app_col_2}
= & \frac{(1-\cos(\alpha)) \dot \gamma m}{3 a h N_c} \lla N_{sp} N_{sc} \rra,
\end{align}
with Eq. (\ref{app_cm_vel_3}). The number of particles $N_{sc}$
and $N_{sp}$ are binomially distributed, \cite{kiku:03} which
yields the average $\lla N_{sp} N_{sc} \rra = N_c (N_c-1) \Delta z
(1 -\Delta z/a)/a $. $\Delta z/a$ and $1-\Delta z/a$ are the
probabilities to find a particle in one of the respective parts of
a collision cell. The average over the random shift $\Delta z$
($0< \Delta z < a$) yields
\begin{align} \label{app_col_5}
\lla \Delta z \left(1- \frac{\Delta z}{a} \right) \rra = \frac{1}{a} \int_0^a \Delta z
\left(1- \frac{\Delta z}{a} \right) \  d\Delta z = \frac{a}{6}.
\end{align}
Thus,
\begin{align} \label{app_col_6}
\sigma_{xz}^{\mathrm{col}} =
\frac{m \dot \gamma}{18 a h} (1-\cos(\alpha))\left(N_c - 1 \right)
\end{align}
and the collisional viscosity is given by Eq. (\ref{visc_col})\\

\section{Surface kinetic stress tensor}

The kinetic contribution to the stress tensor (\ref{stress_split})
at a wall is given by
\begin{align} \label{app_kin_1}
\sigma_{xz}^{\mathrm{kin}} =\frac{L}{V h} \lla \sum_{i=1}^{N_s} \Delta p_{ix}^w \rra.
\end{align}
The average contains the information about the number of particles
colliding with a wall in a streaming step, which can be determined
by applying kinetic theory. The number of particles in a volume
element $Vdz d{\bm v} /L$ of the one-particle phase-space is given
by $dN = N_s P({\bf v}) dz d{\bm v}/L$, with $P({\bm v})$ the
velocity distribution function. Hence, Eq. (\ref{app_kin_1}) can
be reformulated as
\begin{align} \label{app_kin_2}
\sigma_{xz}^{\mathrm{kin}} =\frac{N_c}{ h} \int  \int_{L/2-hv_z}^{L/2}
\Delta p_x^w P({\bm v}) \ dz d{\bm v}
\end{align}
for the surface at $z_w = L/2$. Only particles with velocities
$v_zh > 0$ are able to reach the surface in the collision time
interval $h$. By substitution of the momentum and with $v_x' = v_x
- \dot \gamma z$, we obtain
\begin{align} \nonumber
\sigma_{xz}^{\mathrm{kin}} = & -\frac{2 m N_c}{ h} \int  \int_{L/2-hv_z}^{L/2}
(v_x -u) P({\bm v}) \ dz d{\bm v} \\ \nonumber = &
 -\frac{2 m N_c}{ h} \int  \int_{L/2-hv_z}^{L/2}
(v_x' -u + \dot \gamma z) P({\bm v}) \ dz d{\bm v}\\ \label{app_kin_3} = &
 -2 m N_c \int \left( v_x' v_z - \frac{\dot \gamma h}{2} v_z^2 \right) P({\bm v})
  \  d{\bm v} .
\end{align}
Extending the velocity integration from $0<v_z< \infty$ to
$-\infty < v_z < \infty$ yields
\begin{align} \label{app_kin_4}
\sigma_{xz}^{\mathrm{kin}} =
 - m N_c \left( \lla v_x' v_z \rra - \frac{\dot \gamma h}{2} \lla v_z^2 \rra\right) .
\end{align}
The velocities in this equation are the velocities after
collision. With Eq. (\ref{rel_stream}), the above stationary state
correlation function can be replaced by the correlation function
of the velocities after streaming, which gives
\begin{align} \label{app_kin_5}
\sigma_{xz}^{\mathrm{kin}} =
 - m N_c \left( \lla \hat v_x' \hat v_z \rra + \frac{\dot \gamma h}{2} \lla \hat v_z^2 \rra\right) .
\end{align}
This equation agrees with the corresponding expression in Eq.
(\ref{stress_in_shear}) and we therefore obtain the same
analytical expression for the kinetic viscosity as for the
periodic system, namely Eq. (\ref{visc_kin}).

\end{document}